\documentclass[prd,twocolumn,amsmath,nobibnotes,nofootinbib,superscriptaddress]{revtex4}

\usepackage{bm} \usepackage{graphicx}
\usepackage{epstopdf}
\def\VEV#1{{\left\langle #1 \right\rangle}}
\def\lsim{\mathrel{\rlap{\lower4pt\hbox{\hskip1pt$\sim$}}
    \raise1pt\hbox{$<$}}}         
\def\gsim{\mathrel{\rlap{\lower4pt\hbox{\hskip1pt$\sim$}}
    \raise1pt\hbox{$>$}}}         

\begin{document}

\title{Dynamical and Gravitational Instability of
Oscillating-Field Dark Energy and Dark Matter}
\date{\today}

\author{Matthew C.~Johnson}
\email{mjohnson@theory.caltech.edu}
\affiliation{California Institute of Technology, Mail Code 130-33, Pasadena, CA 91125, USA}
\author{Marc Kamionkowski}
\email{kamion@tapir.caltech.edu}
\affiliation{California Institute of Technology, Mail Code 130-33, Pasadena, CA 91125, USA}

\begin{abstract}
Coherent oscillations of a scalar field can mimic the behavior
of a perfect fluid with an equation-of-state parameter
determined by the properties of the potential, possibly driving
accelerated expansion in the early Universe (inflation) and/or
in the Universe today (dark energy) or behaving as dark
matter. We consider the growth of inhomogeneities in such a
field, mapping the problem to that of two coupled anharmonic
oscillators. We provide a simple physical argument that
oscillating fields with a negative equation-of-state parameter
possess a large-scale dynamical instability to growth of
inhomogeneities. This instability renders these models unsuitable for
explaining cosmic acceleration. We then consider the gravitational instability of oscillating fields in potentials that are close to, but not precisely, harmonic.  We use these results to show
that if axions make up the dark matter, then the small-scale cutoff in the matter
power spectrum is around $10^{-15}\,M_\oplus$. 

\end{abstract}

\maketitle

\section{Introduction}

Observational evidence for inflation, for accelerated
expansion in the current Universe \cite{supernovasearches}, and
for nonbaryonic dark matter has motivated the search for
exotic forms of energy.  Scalar fields have been thoroughly
investigated for both pressureless and negative-pressure matter.
If the kinetic and potential energies of
the scalar field are nearly equal, then the field behaves
as pressureless matter. The most important example is axion
dark matter, which can be described as coherent oscillations of
the axion field in a nearly harmonic potential. Alternatively, if the ratio of potential energy to kinetic energy is sufficiently large, cosmic
acceleration can be induced.
Quintessence achieves this with a rolling scalar field, the rolling
slowed by Hubble friction \cite{quintessence}.  Spintessence
\cite{Boyle:2001du,Gu:2001tr}
achieves this with a complex scalar
field rotating in an internal $U(1)$ symmetric potential; here,
the centripetal acceleration, rather than Hubble friction,
prevents the scalar field from falling directly to its minimum. A coherently oscillating scalar field can also drive cosmic acceleration.

Coherent oscillations in a harmonic potential behave as
nonrelativistic matter, but oscillations in a more general
potential can mimic a perfect fluid with an arbitrary
equation-of-state parameter. For example, in the presence of a
power-law potential $V(\phi) \propto
|\phi|^n$, one finds that $w=(n-2)/(n+2)$~\cite{Turner:1983he}. The power-law index $n$ determines,
through the virial theorem, how the kinetic- and potential-energy densities are apportioned over one oscillation
cycle. When $n<1$, we have $w<-1/3$, illustrating that coherent oscillations of a scalar field might drive a period of inflation in the early
Universe \cite{Damour:1997cb,Liddle:1998pz,Lee:1999pta,Cardenas:1999cw,Sami:2001zd,Sahni:2001qp,Liddle:1998xm,Sami:2003my} or provide a candidate for dark energy~\cite{Sahni:1999qe,Masso:2005zg,Gu:2007be,Hsu:2003ux,Dutta:2008px,Sami:2002se}.

The purpose of this paper is to investigate the dynamical (i.e., those arising from the scalar-field
dynamics) and gravitational instabilities to the growth of
inhomogeneities that may arise in oscillating-field matter. 
The presence of such inhomogeneities are important for 
determining the viability of these models for describing inflation, 
dark energy, and dark matter. 

The question of dynamical stability of oscillating-field dark
energy has been analyzed, in the context of inflation, in
Refs.~\cite{Tsujikawa:2000kw,Taruya:1998cz}, concluding that
oscillating fields that give rise to accelerated expansion are
indeed unstable to the growth of inhomogeneities.  This prior
work analyzed the equations of motion, and it focused on the
resonant growth of perturbations on small scales.
Ref.~\cite{Kasuya:2002zs} considered further the possible
nonlinear evolution of the field.
Refs.~\cite{Damour:1997cb,Sahni:1999qe,Masso:2005zg} speculated that
accelerating oscillating potentials may have a large-scale
instability, but they did not carry out a full instability analysis. In spintessence, an analysis of the linearized equations of motion
for scalar-field perturbations about the homogeneous solution
shows that many models driving cosmic acceleration (for a
power-law potential, large-scale instabilities set in for $n<2$)
are dynamically unstable~\cite{Boyle:2001du,Kasuya:2001pr}.

In addition, prior work~\cite{Khlopov:1985jw,Hu:2000ke} has shown that
gravitational instabilities of oscillations in a harmonic potential, suitable for oscillating-field 
dark matter, are suppressed on sufficiently small scales. This scale determines the 
small-scale cutoff in the matter power spectrum, and has potentially interesting implications for cosmology.

Our principle new contribution to the dynamical stability of
oscillating-field matter is a simple physical picture of
the criterion for stability at large scales. We show that the
perturbed scalar-field equation of motion is identical to that of two
coupled anharmonic oscillators.  This picture then allows us to
provide a simple understanding of why oscillating potentials with
negative pressure should be unstable at large scales, while
those with positive pressure should have large-scale stability.
We then verify these conclusions analytically for potentials
that are nearly harmonic (i.e., nearly pressureless) and
numerically for a broader range of potentials. Our
gravitational analysis generalizes prior work
\cite{Khlopov:1985jw,Hu:2000ke} by considering potentials that
are nearly, but not precisely, harmonic.

Below, we first review (in Section II) the homogeneous evolution
of an oscillating scalar field.  We then provide in Section III
a heuristic discussion of the dynamical instability, beginning
first with an explanation for the origin of the dynamical
instability in spintessence models.  Section IV shows analytically
that for nearly harmonic potentials, instability occurs for
negative-pressure potentials and stability occurs for
positive-pressure potentials.  Section V verifies this
conclusion numerically for more general potentials.  Section VI
includes gravity in the analysis and works out the
gravitational-instability scale for nearly harmonic potentials.
We discuss here the application to axion dark-matter models,
working out the small-scale cutoff in the matter power spectrum.  Section VII presents some
concluding remarks.

\section{Homogeneous Evolution}

Consider a scalar field with potential
$V(\phi)=V(-\phi)$ minimized at $V(\phi=0)=0$.  The equation of
motion for the scalar field is 
\begin{equation}\label{eq:bgrndeqn}
\ddot \phi + V'(\phi) = 0,
\end{equation}
where the dot denotes derivative with respect to time $t$ and the
prime denotes a derivative with respect to the scalar field $\phi$.
The scalar field will undergo periodic motion in this potential,
with a period
\begin{equation}\label{eq-period}
     T (\phi_0) \equiv \frac{2 \pi}{\omega(\phi_0)} = 4
     \int_{0}^{\phi_0}\, d\phi\, \left[V(\phi_0) - V(\phi) \right]^{-1/2},
\end{equation}
that depends, most generally, on the scalar-field amplitude
$\phi_0$.  The period can also be written in terms of the action
\cite{Masso:2005zg}
\begin{equation}
   J = 4 \int_{0}^{\phi_0}\, d \phi \, \sqrt{2 \left[V(\phi_0) -
   V(\phi) \right]},
\label{eqn:action}
\end{equation}
as $T=dJ/dV_0$, where $V_0 = V(\phi_0)$.
  
For $V(\phi) \propto |\phi|^n$, the angular frequency
$\omega(\phi_0) \propto |\phi_0|^{(1/2)-(1/n)}$.  Thus, for
$n=2$, the frequency is amplitude-independent. For $n>2$, the
frequency increases with amplitude, and for $n<2$, the frequency
decreases with amplitude.  
  
On timescales long compared with the oscillation period, the
scalar-field oscillations behave like a perfect fluid with
energy density $\rho$ and pressure $p$,
\begin{eqnarray}
   \label{eq-rho} \rho &\equiv& \VEV{\dot{\phi}^2 /2+V}, \\
   \label{eq-p} p &\equiv& \VEV{\dot{\phi}^2 /2 - V},
\end{eqnarray}
where the angle brackets denote a time average.  The virial
theorem tells us that $\VEV{\dot\phi^2/2}=(1/2)(1+w) V_0$ and
$\VEV{V(\phi)}=(1/2)(1-w) V_0$, where $V_0\equiv V(\phi_0)$ and
$w\equiv p/\rho=(n-2)/(n+2)$ is the equation-of-state parameter.
These results can be summarized in terms of the action by
\cite{Masso:2005zg}
\begin{equation}
     w = \frac{J}{V_0} \frac{1}{\left( \frac{dJ}{dV_0} \right) }
     - 1.
\label{eqn:wequation}
\end{equation}
Differentiating the relation $T=dJ/dV_0$ and re-arranging
algebraically, we obtain an expression for the variation of the
frequency with field amplitude,
\begin{equation}
\frac{d \omega}{d \phi_0} = \frac{\omega^2}{2 \pi (1+w)^2} \frac{dV (\phi_0)}{d \phi} \frac{J}{V_0} \left[ \frac{dw}{dV_0} + \frac{w}{V_0} \right].
\label{eqn:dTdV0}
\end{equation}
Thus, the frequency increases with amplitude unless
$V_0 dw/dV_0 + w < 0$. 

If $w$ is independent of $\phi_0$, as for
power-law potentials, then the frequency will 
decrease with amplitude when $w < 0$ and increase with amplitude
when $w>0$.  If $w$ is allowed to change with $\phi_0$, then we
can have $w<0$ and $d\omega/d\phi_0 \leq 0$, but only temporarily.

\begin{figure}
\includegraphics[width=7.5cm]{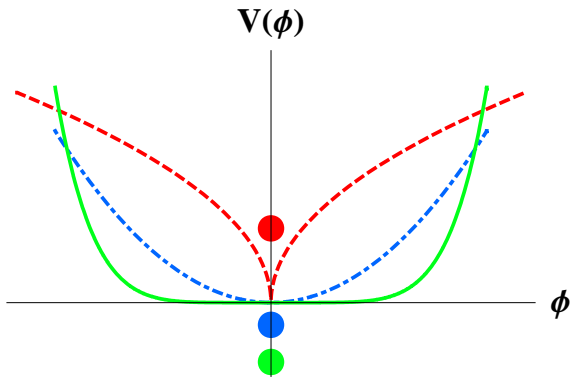}
\caption{Examples of potentials. The red dashed potential
produce an equation of state corresponding to accelerated
expansion ($w < -1/3$). The average intercept for some amplitude
is shown as the red dot at $V>0$. The harmonic potential, which
produces an equation-of-state parameter $w=0$, is the blue
dot-dashed curve, and its average intercept the blue dot at
$V<0$. Oscillations in the solid green potential produce an
equation-of-state parameter $w > 0$, with an average intercept at the
green dot at $V<0$. \label{fig-expots}}
\end{figure}

There is also a nice geometric interpretation of the value of
the equation-of-state parameter~\cite{Damour:1997cb}. Using
energy conservation and the field equation, we can write
\begin{equation}\label{eq-accelcondition}
     \VEV{V - \phi \frac{dV}{d \phi}} = - \left(1+3 w\right)
     \frac{V_0}{2}.
\end{equation}
The left-hand side is the average of the intercept of the tangent to the
potential, and is in general positive for convex (about $\phi =
0$) potentials and negative for non-convex potentials. A few
simple example potentials are shown in Fig.~\ref{fig-expots} to
illustrate this, although it is easy to imagine potentials with
more complicated features. In general, to produce accelerated
expansion, there must be a relatively flat region of the
potential with positive energy somewhere along the oscillation cycle, according with
the intuition that the energy density must be potential
dominated in order to produce accelerated expansion. 

We also note that the sound speed (squared) $c_s^2$ is in
general different than the equation-of-state parameter $w$,
since the sound speed is given by \cite{Masso:2005zg}
\begin{equation}
  c_s^2 = \frac{d (w \rho) }{\rho} = V_0 \frac{dw}{dV_0} + w .
\label{eqn:soundspeed}
\end{equation}
We thus see that the sign of the sound speed is the same as the
sign of $dT/dV_0$.

\section{Heuristic Discussion of Dynamical Instability}

\subsection{Preview: the spintessence instability}

We begin by considering the growth of perturbations in spintessence.
These models introduce a complex scalar field $\phi$ with a
$U(1)$-symmetric potential $V(\phi)=V(|\phi|)$.  The scalar
field moves in a circular orbit in the potential at some
constant amplitude $|\phi|$.  There is kinetic energy associated with the
spinning and potential energy associated with the displacement
of the field from the minimum; the balance between the two
is such that the equation-of-state parameter is
$w=[|\phi| V'(|\phi|)-2V(|\phi|)]/[|\phi|V'(|\phi|)+2V(|\phi|)]$.
For example, if $V(|\phi|) \propto |\phi|^n$, then
$w=(n-2)/(n+2)$, matching the result for an oscillating real field.

Ref.~\cite{Boyle:2001du} showed (see also
Refs.~\cite{Qballs}) that this coherently spinning
field remains stable (neglecting gravity) to small perturbations
if $V'(|\phi|)/|\phi| - V''(|\phi|)<0$ and unstable, at
sufficiently long wavelengths, if $V'(|\phi|)/|\phi| -
V''(|\phi|)>0$.  

This result can be understood simply. The
dynamics of perturbations in the linear regime are identical
to the evolution of two particles undergoing circular orbits in a two-dimensional circularly symmetric potential $V(r)$;
the gradient-energy density in the scalar field acts as a spring
of force constant $k^2$ (where $k$ is the Fourier wavenumber of
the perturbation) connecting the two particles. The radii of the two
orbits differ initially by only a tiny amount. In the absence
of any coupling between the two particles, they will evolve
independently, spinning at two slightly different angular
frequencies and each staying at its original amplitude. If
$V'(|\phi|)/|\phi| - V''(|\phi|)<0$, then the
particle at larger $r$ will run slightly ahead, and if
$V'(|\phi|)/|\phi| - V''(|\phi|)<0$, then it will run slightly
behind.  

Now suppose there is a very strong spring that attaches the two particles.  In this
case, the two particles will be bound to each other, and there
will be no growth of perturbations. If there
is a very weak spring, there will be energy transfer
between the two particles on a timescale longer than the period of oscillation.  
In any confining potential, the angular momentum (per unit mass) increases as $r$ increases.  
If the particle at large $r$ runs ahead, as it will when $V'(|\phi|)/|\phi| -
V''(|\phi|)<0$, then it pulls the inner particle forward, donating some
of its angular momentum in the process.  In this way, the
inner particle evolves to a slightly larger $r$ and the outer to
a slightly lower $r$, decreasing their separation, and implying that the perturbation is stable. If, however,
the particle at larger $r$ runs behind, as it will for $V'(|\phi|)/|\phi| - V''(|\phi|)<0$, then it pulls on the
inner particle, taking angular momentum from it.  The
inner particle must then evolve to a smaller-$r$ orbit and the outer
particle to a larger-$r$ orbit.  In this way, the small
initial separation between the particles is amplified, and the
perturbations are unstable. Readers familiar with
accretion-disk physics will recognize this instability as the
source of angular-momentum transport in disks (see, e.g.,
Ref.~\cite{accretion}) .  

For $V(|\phi|) \propto |\phi|^n$, the instability sets in for $n<2$.  It is
thus concluded that power-law spintessence potentials with
negative pressure are subject to this instability and are
therefore unsuitable as dark-energy candidates.

\subsection{The oscillating-field instability}\label{sec:qualinstab}

A similar argument can be applied to the growth of
perturbations in oscillating-field matter. Adding a perturbation $\delta\phi(\vec x,t)$ to
the homogeneous solution for the scalar field, the linear equation of motion for the perturbation obtained from Eq.~(\ref{eq:bgrndeqn}) is
\begin{equation}
     \ddot{\delta\phi} -\nabla^2\delta\phi +
     V''(\phi)\delta\phi=0.
\end{equation}
In linear theory, each Fourier mode of the density field
$\delta\phi_{\vec k}$ evolves independently and satisfies an
equation (suppressing the $\vec k$ subscript),
\begin{equation}
     \ddot{\delta\phi} + k^2 \delta\phi+V''(\phi)\delta\phi=0.
\label{eq:lineareom}
\end{equation}
This equation of motion is identical to that for the separation
between two particles connected by a spring of force constant $k^2$ moving in a potential $V(\phi)$ with a
separation $\delta\phi\equiv \phi_2 - \phi_1 \ll \phi_1,\phi_2$.

In this picture, particle 1 is released at rest from some initial height $\phi_1 \sim \phi_0$ (where we will later identify $\phi_0$ as the amplitude of the background oscillations), and particle 2 is released slightly higher at $\phi_2 = \phi_1 + \delta \phi$. Because the frequency of oscillation will in general be amplitude dependent, if there is no spring connecting the two masses, then each oscillates at its own frequency. If
$\omega'(\phi_0)>0$, then the higher-amplitude mass (particle 2)
runs ahead of the lower-amplitude mass (particle 1), and {\it
vice versa} if $\omega'(\phi_0)<0$.  

If the spring connecting the two masses is extremely strong, then the two masses oscillate
together. However, if the spring is weak, then
energy can be exchanged between the two particles. If
$\omega'(\phi_0)>0$, then particle 2 runs ahead, pulls on particle
1, and consequently donates some of its energy.  Particle 1 then
moves to a higher-amplitude orbit, particle 2 to a lower one,
and the separation between them diminishes. Such perturbations
are stable. If $\omega'(\phi_0)<0$, then particle 1
runs ahead and donates energy to particle 2. Particle 1 thus
moves to a lower-amplitude orbit, and particle 2 to a higher
orbit. In this case, the initial separation is amplified, and
perturbations are unstable.  

\begin{figure*}
\includegraphics[width=6cm]{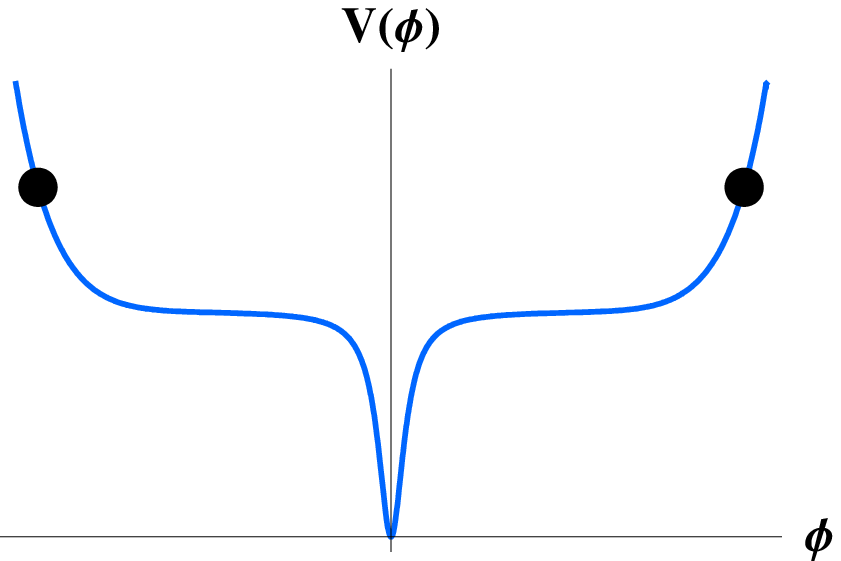}
\includegraphics[width=6cm]{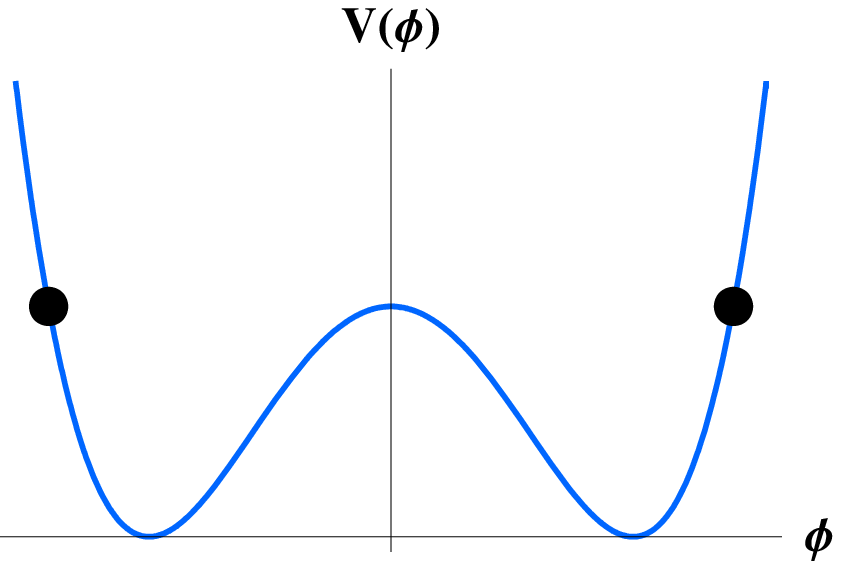}
\caption{Two examples of a potential that can produce
     accelerated expansion and have stability on small
     scales---but only with oscillation amplitudes near those
     indicated by the dots. \label{fig-noinstab}}
\end{figure*}

We conclude that the sign of $\omega'(\phi_0)$ provides a
criterion for stability. This can be obtained from
Eq.~(\ref{eqn:dTdV0}) as $sign(\omega') = sign(V_0 dw/dV_0 +
w)$. As discussed above, power-law potentials with negative
pressure have $\omega'(\phi_0)<0$, and so will develop
large-scale (small $k$) instabilities, rendering these models
unsuitable for accounting for dark energy. Using the geometrical
condition Eq.~(\ref{eq-accelcondition}) as guidance, it is
possible to find potentials that produce
cosmological acceleration and do not develop large-scale
instabilities, but only for a small range of amplitudes. The
potential in the left cell of Fig.~\ref{fig-noinstab}, which is
of the form suggested by Ref.~\cite{Masso:2005zg}, will exhibit
long-term cosmic acceleration, but only short-term stability as
the amplitude of oscillation inevitably decays. The potential in
the right cell of Fig.~\ref{fig-noinstab} will exhibit long-term
stability, but only short-term cosmic acceleration. The analysis
of the viability of these models is somewhat more involved,
although a significant and perhaps unnatural tuning of both the
initial conditions and the potential seem necessary.

Of course, care must be taken with the above arguments, as the
energy exchange between the particles must be considered
throughout the particles' orbits, and not just at the outset of
their motion.  Nevertheless, as we shall discuss further below,
we have been able to verify analytically that perturbations in
nearly-harmonic potentials (to be defined more precisely below)
with $\omega'(\phi_0)<0$ are unstable, while those with
$\omega'(\phi_0)>0$ are stable.  We will then discuss numerical
results that support these conclusions. 

\section{Quantitative Analysis of Dynamical Instability}

We now return to Eq.~(\ref{eq:lineareom}) to discuss the
quantitative behavior of perturbations. In a harmonic potential,
$V''=\omega^2$ is constant, and $\delta\phi$
oscillates with fixed amplitude for all $k^2$. In the most
general potential, $V''(\phi)$ oscillates---although not
necessarily sinusoidally---with an oscillation frequency
$2 \omega$, and so the perturbation equation,
Eq.~(\ref{eq:lineareom}) is that of a harmonic oscillator with an
oscillating mass.

The solutions to Eq.~(\ref{eq:lineareom}), as well as the
issue of their stability, is the subject of Floquet
theory.\footnote{If we replace time $t$ by a position $x$, our
equation becomes the Schrodinger equation for a particle in a
periodic potential, the solutions of which are described by
Bloch's theorem.}  There is no simple stability condition in the
most general case, but
we can obtain an analytic solution for potentials that are close
to harmonic.  To begin, consider the quartic potential $V(\phi)
=(1/2)m^2\phi^2 +(\lambda/4) \phi^4$, working to linear order in
$\lambda$ in the limit $\lambda \rightarrow0$.  In this case,
$V''(\phi) = m^2 + 3\lambda \phi^2$.  If the oscillation
amplitude is $\phi_0$, the oscillation frequency is
$\omega^2=m^2[1+(3/4) \lambda \phi_0^2/m^2]$ and the homogeneous
oscillation is nearly sinusoidal: $\phi(t) =\phi_0 \cos \omega
t$.  With the trigonometric identity, $\cos^2 x = (1/2)(1+\cos
2x)$, the perturbation equation is, to linear order in
$\lambda$, then
\begin{equation}
     \ddot\delta\phi + \left\{ [k^2 + m^2 + (3/2) \lambda
     \phi_0^2] + (3/2) \lambda \phi_0^2 \cos 2\omega t \right\}
     \delta\phi=0.
\end{equation}
Defining $z=\omega t$, this is identified as the Mathieu
equation,
\begin{equation}
     \frac{d^2}{dz^2} \delta\phi + [a- 2q \cos 2z]\delta \phi=0,
\end{equation}
with 
\begin{equation}
     a=\left(1+\frac{k^2}{m^2}\right) \left(1-\frac{3}{4}
     \frac{\lambda \phi_0^2}{m^2}\right) + \frac{3}{2}
     \frac{\lambda \phi_0^2}{m^2}, \quad q =
     -\frac{3}{4}\frac{\lambda \phi_0^2}{m^2}.
\end{equation}
If we consider only
long-wavelength (i.e., $k \ll m$) fluctuations and the limit $\lambda
\phi_0^2 \ll m^2$, then $a\simeq 1$ and $q\ll 1$.  In this
regime, $\delta \phi$ oscillates rapidly with frequency
$\omega t$ with an amplitude that varies as $e^{\pm \Omega t}$
with $\Omega \simeq \sqrt{q^2-(1-a)^2}\omega$.  Thus, the
condition for instability is $1-|q|<a<1+|q|$ (as derived in
Section \ref{sec:gravinstability} below.)

Thus, if $\lambda>0$, then the solutions are stable for all $k^2>0$.  For
$\lambda<0$, the solutions are stable only if $k^2>k_J^2 =
-(3/2)\lambda \phi_0^2$.  Recalling that the oscillation
frequency is $\omega^2 \simeq m^2 +3\lambda\phi_0^2$, we see
that there is instability, on sufficiently large scales $k^{-1}$,
if $\omega'(\phi_0)<0$, while the perturbations are stable for
all $k$ if $\omega'(\phi_0)>0$.  

We can also generalize to other potentials that are close to
harmonic, by which we mean that the time
dependence of $V''[\phi(t)]$ can be approximated as $V''(t)
\approx V^{\prime\prime}_0 +  V^{\prime\prime}_2 \cos 2\omega t$
with constant $V^{\prime\prime}_0$ and $V^{\prime\prime}_2$.
The Jeans wavenumber $k_J$ is then given by
\begin{equation}
     k_J^2 = \frac{1}{2} | V^{\prime\prime}_2| + \omega^2 -
     V^{\prime\prime}_0.
\label{eq:approxkJ}
\end{equation}
It can be checked numerically that for the power-law
potentials $V(\phi) \propto |\phi|^n$ with $n\simeq2$, these
relations imply stability for $n>2$ and instability for $n<2$,
as our heuristic arguments suggest.  

Of course, the types of potentials that can drive cosmic
acceleration are far from harmonic (for the power-law potential,
an index $n<1$ is required), and so the analysis presented above
will no longer be valid. However, we expect no qualitative
difference, and to confirm this we investigate the stability of
oscillations in more complicated potentials numerically in the
next Section.

\section{Numerical Results on Dynamical Instability}\label{numerical}

We will numerically determine the stability of oscillations in
the presence of three different representative classes of
potential in this section: power-law potentials with arbitrary
index, temporarily stable potentials exhibiting cosmic
acceleration (as in the left cell of Fig.~\ref{fig-noinstab}),
and stable potentials exhibiting temporary cosmic acceleration
(as in the right cell of Fig.~\ref{fig-noinstab}). In each case,
we write the potential in the form $V(\phi) = \mu^4 v(\phi /
\phi_0)$, and define the dimensionless variables,
\begin{equation}
     x = \phi / \phi_0, \ \ \tau = \frac{\mu^2}{\phi_0} t, \ \
     \kappa = \frac{\phi_0}{\mu^2} k, \ \ y = \frac{\delta
     \phi}{A},
\end{equation}
where $A$ is an appropriately defined constant characterizing the amplitude of the perturbation. The equations of motion then become
\begin{eqnarray}
     0 & =& \frac{d^2 x}{d\tau^2} + \frac{dv}{dx},   \nonumber \\
     0 &=& \frac{d^2 y}{d \tau^2} + \left[ \kappa^2 +
     \frac{d^2v}{dx^2} \right] y.
\label{eq:dimlesseqns}
\end{eqnarray}

To determine the regions of stability, we can form the
fundamental solution matrix (for a more detailed discussion of
determining the stability of Hill's equation, see, e.g.,
Ref.~\cite{Magnus:1966fk})
\begin{equation}
     C= \left(
     \begin{array}{cc}
     y_{1} (T/2) & y_{2} (T/2) \\
     dy_1/d\tau (T/2) & dy_2/d\tau (T/2) 
     \end{array}
     \right),
\end{equation}
where $y_1$ is the solution generated from the initial
conditions $\left\{ y(0)=1,dy/d\tau (0)=0 \right\}$, and $y_2$ is
the solution generated from the initial conditions $\left\{
y(0)=0,dy/d\tau(0)=1 \right\}$. The stability of a solution can
be determined by the eigenvalues of $C$ (noting that $det(C) =
1$)
\begin{equation}
     \Lambda_{\pm} = \frac{tr(C) \pm \sqrt{tr(C)^2 - 4}}{2}.
\end{equation}
Instability results when either eigenvalue has modulus greater
than unity. Since $\Lambda_+ \Lambda_- = 1$ (from the
determinant condition), if the roots are both real, then
solutions are unstable. This occurs when $|tr(C)| > 2$. It also
follows that on the boundary between stability and instability,
$y_1(t)$ is a function of period $T/2$ or $T$ (either the same
period as $v''$ or $x$). Because the system of equations we are
studying are invariant under time shifts by a period of
oscillation, from $y_{1} (0) = dy_2/d\tau (0) = 1$, we will have
$y_{1} (T/2) = dy_2/d\tau (T/2)$. The stability criterion can in
this case be phrased as $|y_{1} (T/2)| < 1$, and provides
an easy way to numerically determine stability or instability.

\begin{figure}
\includegraphics[width=8.5cm]{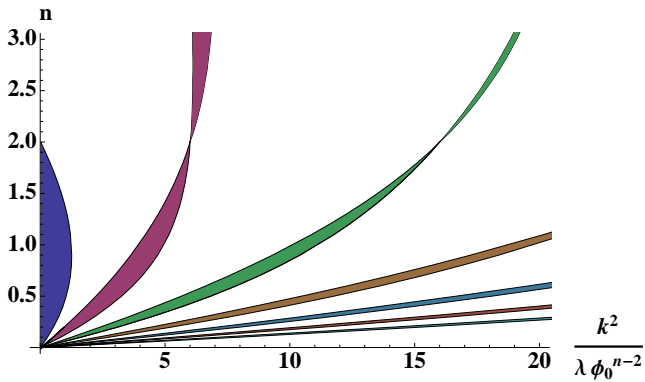}
\caption{The first seven bands of instability for oscillations
     in the presence of a power-law potential.} 
\label{fig:powerlawbands}
\end{figure}

We will first consider power-law potentials 
\begin{equation}
     V(|\phi|) = \lambda |\phi|^n.
\end{equation}
The singularity at the origin can be regulated by introducing a
numerically small smoothing parameter $c$, such that $V(|\phi|)
= \lambda (c+\phi^2)^{n/2}$. We have checked that the numerics
are insensitive to the value of this parameter. Numerically
integrating Eq.~(\ref{eq:dimlesseqns}), and solving for the roots
of $|y_{1} (T/2)| - 1$, we can find the boundaries of
stability. Plotted in Fig.~\ref{fig:powerlawbands} are the seven
lowest bands of instability (the shaded regions of the Figure)
for power-law potentials with $n<3$. It can be seen that there
is a band of instability encompassing $k=0$ for $n<2$ as
expected from the general arguments given above. In addition, as
$n$ decreases, the higher-order bands of instability migrate
towards lower wavenumber. For power-law potentials with an
equation-of-state parameter $w \sim -1$, the power-law index
will be rather small, and the large number of higher-order bands
of instability at small $k$ might become important in analyzing
the overall stability of such a model. 

\begin{figure}
\includegraphics[width=7.5cm]{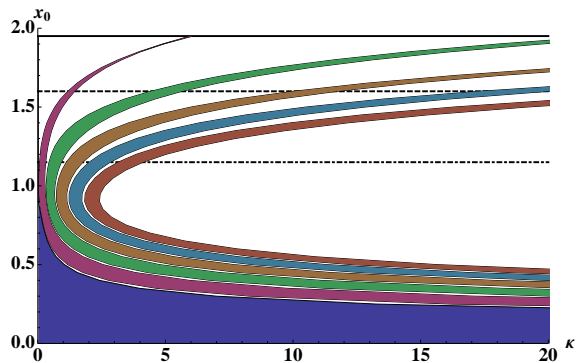}
\caption{The first six bands of instability for oscillations in
     the presence of the potential shown in the left cell of
     Fig.~\protect\ref{fig-noinstab}. The solid line represents an amplitude
     corresponding to $w = -0.4$, dashed line $w=-0.6$ and dot-dashed
     line $w=-0.8$. In this example, $w$ remains less than one, but as the 
     amplitude of oscillations decays a large-scale band of instability develops. } 
\label{fig:tempstable}
\end{figure}

Moving to potentials of the form shown in the left cell of
Fig.~\ref{fig-noinstab}, which we can write as,
\begin{equation}
     V (|\phi|) = \mu^4 \left[ \frac{x^2}{c_1 + x^2} + c_2 x^n \right],
\end{equation}
we chose as an example $c_1=c_2= 0.005$ and $n=10$. Again,
numerically integrating Eq.~(\ref{eq:dimlesseqns}), we plot the
first six bands of instability in
Fig.~\ref{fig:tempstable}. Also over-plotted are the values of
the equation-of-state parameter corresponding to various
amplitudes of the background oscillations. It can be seen that
there are indeed regions where $w < -1/3$ and large-scale
stability exists. However, as the universe evolves and the
amplitude of oscillation decays, more and more bands of
instability will pile up at small $k$. When the amplitude of
oscillation dips below $x_0 \sim 1$ in this model, the solution
will become violently unstable at both small and large k, with
only very small bands of {\em stability}. This will most likely
present a severe challenge to using such models as a candidate
for the dark energy. We have also checked numerically that the
large-scale (encompassing $k=0$) band of instability appears
when the amplitude of oscillation is such that $w'$ changes
sign, supporting, in a more complicated potential, our heuristic
arguments for stability.

\begin{figure}
\includegraphics[width=7.5cm]{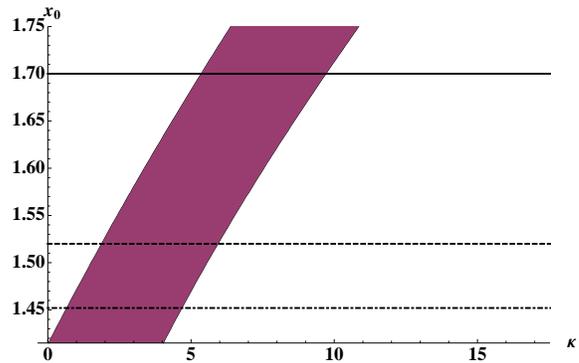}
\caption{The instability band for oscillations in the
     double-well potential shown in the right cell of
     Fig.~\protect\ref{fig-noinstab}. There is just one band of
     instability until $k$ becomes very large. Also plotted are
     lines that indicate the amplitudes at which $w=0.4$,
     $w=0.2$, and $w=0$ (from top to  bottom).  Note that
     $w\rightarrow -1$ as $x\rightarrow \sqrt{2}\simeq 1.41$,
     and so this model features large-scale stability with
     $-1<w<0$.  However, this period of negative pressure is
     short-lived as the Universe expands.}
\label{fig:doublewell}
\end{figure}

Finally, we will treat the case where we expect the potential to
exhibit large-scale stability, but only temporary cosmic
acceleration, as in the right cell of
Fig.~\ref{fig-noinstab}. The potential is given by
\begin{equation}
     V(\phi) = \mu^4 \left( x^2 + c^2 \right)^2,
\end{equation}
where we choose as an example $c = 1$. When the amplitude is
very close to the height of the barrier between the two minima
of this potential (at $x_0=\sqrt{2}$), we expect the
equation-of-state parameter to be $w \sim -1$~\footnote{The
field will loiter near the local maximum, extending the period
of oscillations, and eventually spoiling our approximation that
the period be shorter than a Hubble time; we will ignore this
shortcoming for the moment.}. As shown in
Fig.~\ref{fig:doublewell}, this
potential yields large-scale stability, with the first order
band of instability approaching $k=0$ as the amplitude of
oscillations reaches $x_0 \sim \sqrt{2}$. Also shown are the
values of $w$ for a number of amplitudes.

In all of our examples, we have verified numerically that the
sign of $\omega'$ determines the existence of a band of
instability about $k=0$, as expected from the heuristic
arguments of Sec.~\ref{sec:qualinstab}. We have also illustrated
the existence of higher-order bands of instability that will
influence the development of small-scale inhomogeneities in the
oscillating field. In all examples, the higher bands of instability creep in towards $k=0$ as $w \rightarrow -1$.

\section{Gravitational instability}
\label{sec:gravinstability}

\subsection{General analysis}

We now include gravity in the analysis.  As alluded to above,
perturbations in a perfect fluid with sound speed $c_s$ are
stabilized by pressure gradients for wavenumbers $k>k_J = (4 \pi
G\rho/c_s^2)^{1/2}$, but care must be taken in applying this
result to coherent scalar fields. For example, oscillations in
a harmonic potential have a sound speed $c_s^2 =
d(w\rho)/d\rho=0$, implying that perturbations on all scales will
suffer a gravitational instability.  However, as shown in
Ref.~\cite{Khlopov:1985jw}, perturbations on scales $k>2
\sqrt{\pi G} m^2 \phi_0$ are stabilized.  We therefore extend
the analysis of Ref.~\cite{Khlopov:1985jw} to see how their
result for the scalar-field stability is altered by the presence
of a small anharmonic term in the potential. This is in fact general, 
since, as we will see, strong dynamical instabilities will completely swamp 
the gravitational instability, implying that the inclusion of gravitational effects is 
only relevant for potentials that are close to harmonic (as we have defined above).

To simplify the analysis, we
restrict our attention to small-wavelength Fourier modes, $k \gg
H$, where $H=(8\pi G \rho/3)^{1/2}$ and $\rho=(1/2)\dot\phi^2 +
V(\phi)$.  This allows us to
neglect the expansion and work with a perturbed Minkowski
metric (in the conformal Newtonian gauge),
\begin{equation}
     ds^2= (1+2\Psi) dt^2 + (1-2\Psi)d\vec x^2.
\label{eqn:metric}
\end{equation}
where $\Psi$ is the Newtonian potential.  Perturbations
$\delta\phi$ in the scalar field may now induce perturbations
$\Psi$ to the metric, which then affect the scalar-field
equation of motion, which is now,
\begin{equation}
     \ddot{\delta \phi} + \left[ V''\left(\phi(t)\right)
     -\nabla^2 \right] \delta \phi = 4 \dot \Psi
     \dot{(\delta\phi)} -2 \Psi V'\left(\phi(t) \right).
\label{eqn:withgravity}
\end{equation}
The potential is determined by the Einstein equation (the
Poisson equation),
\begin{equation}
     \nabla^2\Psi = 4 \pi G \left[\dot\phi \dot{(\delta\phi)} +
     V'\left(\phi(t) \right) \delta \phi  - \Psi \dot\phi^2
     \right].
\label{eqn:einstein}
\end{equation}
We now focus on a given Fourier mode of wavenumber $\vec k$, in
which case we replace $\nabla^2 \rightarrow -k^2$ in these
equations, and we neglect the last term on the right-hand side of
Eq.~(\ref{eqn:einstein}), as it is negligible for $k \gg
H$.  

Although the equations have become more complicated with gravity,
they are once once again linear differential equations with
periodic coefficients, and so the solutions are
formally described by Floquet theory.  

\subsection{A quartic correction to a harmonic potential}
We now consider the potential $V(\phi)=(1/2)m^2 \phi^2 +(\lambda/4)
\phi^4$ and suppose that the anharmonic term is small, $\lambda
\phi_0^2 \ll m^2$. Eqs.~(\ref{eqn:withgravity}) and (\ref{eqn:einstein}) can then
be combined into a single second-order differential equation,
\begin{widetext}
\begin{equation}
     \ddot \delta\phi +m
     \gamma \sin 2\omega t \dot \delta \phi + \left[
     k^2+m^2+\frac{3}{2} \lambda \phi_0^2 -\gamma m^2 + \left(
     \frac{3}{2} \lambda \phi_0^2 - \gamma m^2 \right) \cos
     2\omega t \right]\delta \phi=0.
\label{eqn:fullequation}
\end{equation}
\end{widetext}
where $\gamma \equiv G \phi_0^2 / k^2$.
At the lowest-$k$ stability boundary, the solutions are periodic
with frequency $\omega$, and can thus be written $\delta\phi(t) =
a \cos\omega t + b\sin\omega t$.  We then plug this into
Eq.~(\ref{eqn:fullequation}) and demand that the coefficients
of the terms that vary as $\cos\omega t$ and as $\sin \omega t$
vanish. In doing so, we work to linear order in $\lambda$ and
$\gamma$ and neglect terms of $O(\lambda\gamma)$.  We then find
that the boundary between stability and instability occurs for
wavenumbers $k_1$ and $k_2$ given by
\begin{equation}
     k_1^2 = -\frac{3}{2} \lambda \phi_0^2 +\gamma m^2, \qquad k_2^2
     = 0.
\label{eqn:borders}
\end{equation}
A bit more algebra (replacing $a$ and $b$ by slowly varying
functions $a(t) =a_0 \exp(\Omega t)$ and $b(t)=b_0 \exp( \Omega t)$, with
$\Omega \ll \omega$) shows that the instability occurs
for values of $k$ between $k_1$ and $k_2$.

If $\lambda<0$, then $k_1^2 > 0$. In this case, the field is
dynamically unstable, even without gravity (as noted before),
and gravity only serves to increase the instability.  If,
however, $\lambda>0$, gravity induces an instability (noting
that $\gamma\propto k^{-2}$), for $k^2< k_J^2$, with
\begin{equation}
     k_J^2 = -\frac{3}{4} \lambda \phi_0^2 +
     \sqrt{ \left(\frac{3}{4} \lambda \phi_0^2\right)^2 + 8 \pi
     G m^4 \phi_0^2 }.
\label{eqn:kJwithgravity}
\end{equation}
Note that this recovers our earlier result, without gravity, in
the limit that $G\rightarrow0$, and it recovers the result of
Ref.~\cite{Khlopov:1985jw,Hu:2000ke} for $\lambda=0$.

The result for the Jeans scale resembles that
for spintessence (cf., Eq.~(10) in Ref.~\cite{Boyle:2001du})
with the replacement $(1/2)(V'/|\phi|-V'') = -\lambda
|\phi_0^2|$ for spintessence with $-(3/4)\lambda \phi_0^2$ for
the oscillating field, and replacing the $G$ in the spintessence
result by $G/2$. The former replacement occurs because
the sound speed $c_s^2=d(w\rho)/d\rho=(1/2)\lambda(\phi_0^2/m^2)$ for
spintessence differs from that, $c_s^2=(3/8)\lambda (\phi_0^2/m^2)$, for
the oscillating field.  The latter replacement (i.e.,
$G\rightarrow G/2$) occurs because the complex field is
equivalent to two scalar fields.  Note that the result,
Eq.~(\ref{eqn:kJwithgravity}), differs from the spintessence
result also because the spintessence Jeans scale of Eq.~(10) in
Ref.~\cite{Boyle:2001du} places no restrictions on the potential
$V(R)$, while Eq.~(\ref{eqn:kJwithgravity}) here is valid only
in the limit $\lambda \phi_0^2 \ll m^2$.

In the limit that $Gm^4 \phi_0^2 \ll \lambda^2 \phi_0^4$, $k_J^2
\simeq 16 \pi G m^4/3\lambda$, which is equal to $4\pi G/c_s^2$,
the Jeans scale for a perfect fluid.  However, if $Gm^4 \phi_0^2
\gg \lambda^2 \phi_0^4$, then the perfect-fluid description
breaks down, the scalar-field dynamics become important, and 
the Jeans length differs considerably from the perfect-fluid
result, $c_s^2 k_J^2 = 4\pi G\rho$.

\subsection{General result for nearly harmonic potentials}

The result, Eq.~(\ref{eqn:kJwithgravity}), can be generalized to
other potentials that are close to harmonic by replacing $-(3/2)
\lambda \phi_0^2$ in Eq.~(\ref{eqn:kJwithgravity}) by 
$(1/2) |V''_2| + \omega^2 - V''_0$,
\begin{eqnarray}
     k_J^2 = & & \frac{1}{2} \left[ \frac{1}{2} |V''_2| +
     \omega^2 - V''_0 \right.
     \nonumber \\
     &+& \left. \sqrt{\left[\frac{1}{2} |V''_2| + \omega^2 -
     V''_0| \right]^2 + 8 \pi G m^4 \phi_0^2 } \right].
\label{eqn:generalkJ}
\end{eqnarray}

\subsection{Application: Axion dark matter}

We now calculate the small-scale cutoff in the cold-dark-matter power
spectrum under the assumption that the dark matter is composed
of axions with masses $m_a \sim10^{-5}$ eV \cite{axionreviews}.
This cutoff will
determine the masses of the first dark-matter halos to undergo
collapse in the Universe, and it determines the size of the
smallest clumps in the Milky Way halo
\cite{Diemand:2005vz,Kamionkowski:2008vw}.  If weakly-interacting massive
particles (WIMPs) make up the dark matter, then the primordial power
spectrum is suppressed for cosmological mass scales smaller than
$\sim 10^{-4}-10^{2}\, M_\oplus$ (the precise value determined
by the precise WIMP model) \cite{Profumo:2006bv} by kinetic
decoupling of WIMPs.  We will now calculate the analogous
small-scale cutoff if axions make up the dark matter.

Cosmological axions in the $\sim10^{-5}$~eV mass regime are
produced by the misalignment mechanism near
the time of the QCD phase transition, and they may thus be
described by a coherently oscillating scalar field.  The
axion potential has the form $V(\phi)= V_0 [\cos(\phi/f)-1]$.
Today, the axion field oscillates near the minimum of this 
potential where it can be approximated by $V(\phi) \approx
(1/2)m^2 \phi^2 + (\lambda/4) \phi^4$, with $m^2 = V_0/f^2$ and
$\lambda = -(1/6) (m^2/f^2)$.  Here, $f\simeq \Lambda^2/m$ is
the Peccei-Quinn scale, where $\Lambda \sim 100$ MeV is the QCD
scale.  If axions make up the dark matter, then the axion energy
density today is $\rho_q= (1/2)m^2 \phi_0^2 =\Omega_c \rho_c$,
where $\Omega_c\simeq 0.2$ is the cold-dark-matter density and
$\rho_c$ is the critical density, and this relation can be used
to fix $\phi_0$.  Using the scaling $\phi_0^2 \propto
(1+z)^3$, we find also that $(\lambda \phi_0^2)^2 \ll
Gm^4 \phi_0^2$ not only today, but at all redshifts $z\lesssim
3000$ during matter domination.  The axion-axion interactions
implied by the $\lambda \phi^4$ correction to the quadratic
potential therefore have little effect on the Jeans scale, which
is well approximated by the earlier result of
Refs.~\cite{Khlopov:1985jw,Hu:2000ke}.  

The physical Jeans scale is 
given by $k_J^2 = 2\sqrt{2\pi G} m^2 \phi_0$, and
the Jeans mass  then turns out to be $M_J \equiv (4\pi/3)
(\pi/k_J)^3 \rho_m$, where $\rho_m$ is the current matter
density.  Numerically, 
\begin{equation}
     M_J\simeq 1.8 \times 10^{-13}\,
     (m_a/10^{-5}~{\mathrm eV})^{-3/2} (1+z)^{-3/4} M_\oplus.
\label{eqn:axionkJ}
\end{equation}
The collapse redshift for such low-mass halos depends on the
primordial spectral index, but is generally in the range $3000
\gtrsim z \gtrsim 300$, resulting in a small-scale cutoff of
order $10^{-15}\, M_\oplus$ in the primordial power spectrum,
much smaller than that for WIMPs.  Note that there may also be
dynamical instabilities on even smaller scales (related to the
higher-order bands of instability, as in Sec.~\ref{numerical});
see, e.g., the very elegant work of Ref.~\cite{Greene:1998pb}.

\section{Conclusions}

We have considered the dynamical and gravitational amplification
of inhomogeneities in oscillating-field matter.  We provided a
simple physical picture for the origin of instabilities in
oscillating-field matter and in spintessence.  In this picture,
the condition for instability is seen as a condition on the
amplitude dependence of the frequency, a condition that can be
shown to be equivalent to positivity of the sound speed.
This argument was verified analytically for nearly harmonic
potentials and numerically for more general potentials.  We then
included gravity in the analysis, generalizing earlier results
on the gravitational instability of harmonic potentials to
potentials that differ slightly from harmonic.  We used this
result to evaluate the small-scale cutoff in the matter power
spectrum if axions make up the dark matter.  We leave the
implications of this cutoff (roughly $10^{-15}\,M_\oplus$) to
future work.

Our results indicate that potentials that give rise to
accelerated expansion generically suffer dynamical instabilities
to the growth of large-scale inhomogeneities. These instabilities should
render oscillating fields unsuitable to account for dark energy
in the Universe today or for driving inflation in the early
Universe, as both scenarios require the cosmological density to
remain homogeneous for extended periods of time.  It is true
that there may be potentials that drive acceleration and are
stable, but stability and/or acceleration will be only temporary. 
This loophole is thus unlikely to alter our conclusions.

We were able to make progress analytically for both the
dynamical and gravitational instability only for nearly harmonic
potentials.  It would be interesting to see whether analytic
results for the growth of perturbations can be extended to more
general potentials.  We leave the investigation of this question
for future work.

\acknowledgments
We thank D.~Cohen for discussions of Floquet theory, L.~Kofman
for suggesting some useful papers, and C.~Hirata for
discussions and suggestions.  This work was supported by DoE
DE-FG03-92-ER40701 and the Gordon and Betty Moore Foundation.

\end{document}